\documentstyle[preprint,aps]{revtex}
\tightenlines
\def\lsim{\raise0.3ex\hbox{$<$\kern-0.75em\raise-1.1ex\hbox{$\sim$}}}
\def\gsim{\raise0.3ex\hbox{$>$\kern-0.75em\raise-1.1ex\hbox{$\sim$}}}
\begin{document}
\draft
\preprint{KANAZAWA 98-04,
\ March 1998}     
\title{Monopoles and hadron  spectrum in quenched QCD}
\author{
S.Kitahara$^a$,
O.Miyamura$^b$
\footnote{ E-mail address:miyamura@fusion.sci.hiroshima-u.ac.jp},
T.Okude$^c$,
F.Shoji$^c$
and Tsuneo Suzuki$^c$
\footnote{ E-mail address:suzuki@hep.s.kanazawa-u.ac.jp }
}
\address{$^a$
Jumonji University, Niiza, Saitama 352, Japan}
\address{$^b$
Department of Physics, Hiroshima University, Higashi Hiroshima
724, Japan}
\address{$^c$
Department of Physics, Kanazawa University, Kanazawa 920-11, Japan}

\maketitle

\begin{abstract}
We study quenched hadron spectra with Wilson fermion 
in abelian gauge fields extracted by maximal abelian projection 
and in fields induced by monopoles
on $16^3 \times 32 $ and $ 12^3 \times 24$
lattices.
Pion mass squared and 
quark mass defined through the axial Ward identity
satisfy the PCAC relation.  
Gross features of the light 
hadron spectra are almost similar to those in $SU(3)$ gauge fields if 
normalization is made by the square root of the string tension.  
It is also shown that no sizable dynamical mass generation is found 
in the present range of $\kappa$ when the monopole degree of freedom is
 removed from the abelian fields or from the $SU(3)$ gauge fields. 
\end{abstract}

\begin{center}
\end{center}

\newpage

\section{Introduction}
Understanding of color confinement is still an important subject 
of nonperturbative dynamics of quantum chromodynamics(QCD). 
Investigations along 
monopole condensation proposed by t'Hooft\cite{thooft2} have been
developed especially in lattice Monte-Carlo
simulations\cite{suzu93,cosmai,yee,polika,wensley,shiba4,shiba6,ejiri,stack,arasaki,kondo}.
In the scenario, monopoles in abelian projected gauge field 
play a role of electrons in the superconductor.
Color confinement is caused if the monopoles are condensed
\cite{thooft2}.  
Such a scenario is well confirmed 
in lattice compact QED\cite{poly,bank,degrand,peskin,frolich,smit}. 

A key quantity of confinement, 
i.e. the string tension between quark and antiquark, 
has been measured and analyzed along this scenario.
After the abelian projection in the maximal abelian (MA) gauge,
the string tension derived from abelian Wilson loops gives almost the
same value as that of $SU(2)$\cite{suzu93,yotsu}. 
Moreover, 
the monopole contribution to the abelian Wilson loops  alone reproduces
the string tension in 
$T=0$\cite{shiba4}  and $T\neq 0$\cite{ejiri} $SU(2)$ QCD.

The Polyakov loop is an order parameter of finite-temperature
deconfinement transition in 
pure lattice gauge theory.  It is shown that  abelian Polyakov loops written in
terms of  abelian link fields alone play also a role of an order
parameter \cite{hio,yasuta} in $T\neq 0$ $SU(2)$ and $SU(3)$
lattice QCD simulations. Furthermore, an abelian Polyakov loop 
operator can be expressed by a product of contributions from the Dirac
string of monopoles and from photons \cite{suzu95}. 
Only the former part of the abliean Polyakov loop reveals the
confinement-deconfinement transition.
These results are seen also in some unitary gauge fixings.
These results strongly support that the 'tHooft conjecture\cite{thooft2} 
is realized in QCD. 

What happens when quarks are introduced and 
what determines the hadron spectrum?
Quarks and gluons are confined inside hadrons. 
Abelian dominance hypothesis claims that long wave dynamics of QCD 
is well projected on the abelian part of gauge fields after abelian
projection. Then  mass generation of light hadrons 
from scale invariant QCD is expected to
be explained by the abelian and the monopole parts alone.

The expectation has been tested in
the simulations in quenched 
$SU(2)$ QCD with Kogut-Susskind fermions\cite{miya95}.
It has been shown that light pion mode exists 
in the abelian and the monopole gauge fields.
Its mass vanishes in the chiral limit while the
mass of $\rho$ meson remains finite as shown in $SU(2)$ gauge
field. On the other hand, the propagators
behave as a product of free staggered fermions in the photon part. 

The purpose of this paper is to report new results of
hadron spectra in the abelian gauge fields and
the monopole  gauge fields in quenched $SU(3)$ QCD with Wilson fermions.

 In Section \ref{definition} we briefly review
the maximally abelian gauge in
lattice $SU(3)$ QCD and abelian gauge fields from the Dirac string of
monopoles and   photons.
Our results are shown in Section \ref{results}.
We calculate hadron spectra and the quark mass from the axial Ward
identity. 
The PCAC relation is shown to be satisfied.
Finally we examine light hadron spectra in $SU(3)$ gauge field from 
which monopoles are removed.  The pion and the $\rho$ 
meson masses are degenerate irrespective of $\kappa$. The behavior is
similar to those in the free field. 
Summary and  remarks are given
in Section \ref{summary}.

\section{Definition}
\label{definition}
\subsection{Maximally abelian (MA) gauge in $SU(3)$ QCD}
\label{MAgauge}

We adopt usual $SU(3)$ Wilson action for gauge fields.
The MA  gauge is given 
by performing a gauge transformation
$\widetilde{U}(s,\mu) = V(s)U(s,\mu)V^{-1}(s+\hat\mu)$
such that 
\begin{eqnarray}
  R=\sum_{s,\mu}\:(
    |\widetilde{U}_{11}(s,\mu)|^2
   +|\widetilde{U}_{22}(s,\mu)|^2
   +|\widetilde{U}_{33}(s,\mu)|^2 )
\end{eqnarray}
is maximized. Then a matrix 
\begin{eqnarray}
 X(s) =  \sum_{\mu,a}\:[\widetilde{U}(s,\mu)
   \Lambda_a \widetilde{U}^{\dagger}(s,\mu)
  +\widetilde{U}^{\dagger}(s-\hat\mu,\mu)\Lambda_a 
   \widetilde{U}(s-\hat\mu,\mu),\Lambda_a]
   \label{Xdiagonal}
\end{eqnarray}
is diagonalized at all sites on the lattice. Here 

\begin{eqnarray}
 \Lambda_1 = \left( \begin{array}{ccc}
	      1 & 0 & 0 \\
		     0 &-1 & 0 \\
		    0 & 0 & 0    \end{array}\right),
  \Lambda_2 =\left( \begin{array}{ccc}
	      -1 & 0 & 0 \\
		     0 & 0 & 0 \\
		    0 & 0 & 1    \end{array}\right),
   \Lambda_3 =\left( \begin{array}{ccc}
	       0 & 0 & 0 \\
		      0 & 1 & 0 \\
		     0 & 0 &-1     \end{array}\right).
\end{eqnarray}
After the gauge fixing is over,
we can extract an abelian link field\cite{schierholz}
\begin{eqnarray}
   \widetilde{U}(s,\mu) = C(s,\mu)u(s,\mu),
\end{eqnarray}
where
\begin{eqnarray}
  u(s,\mu)            & = &
               \mbox{diag}(e^{i\theta^{(1)}(s,\mu)}
               ,e^{i\theta^{(2)}(s,\mu)},e^{i\theta^{(3)}(s,\mu)}) \\
  \theta^{(i)}(s,\mu) & = & \mbox{arg}(\widetilde{U}_{ii}(s,\mu))
               -\frac{1}{3}\phi(s,\mu)                      \\
  \phi(s,\mu) & = & \left[\sum_{i}\mbox{arg}(\widetilde{U}_{ii}(s,\mu))
                 \,\right]_{\mbox{{\rm mod} $2\pi$}}\ \
   \in [-\pi, \pi).
\end{eqnarray}
$u(s,\mu)$ is a diagonal abelian gauge field and $C(s,\mu)$ is 
a charged matter field.

\subsection{Abelian gauge fields from the Dirac string and photon}
\label{abeliangf}

A plaquette variable is given by
$f^{(i)}_{\mu\nu}(s)= \partial_{\mu}\theta^{(i)}_{\nu}(s) -
\partial_{\nu}\theta^{(i)}_{\mu}(s)$. Hence
\begin{eqnarray}
  \theta^{(i)}_\mu (s)= \!\!-\!\! \sum_{s'} D(s-s')
  [\partial'_{\nu}f^{(i)}_{\nu \mu}(s')+
  \partial_\mu (\partial'_{\nu}\theta^{(i)}_{\nu}(s'))],
\end{eqnarray} 
where $D(s-s')$ is the lattice Coulomb propagator. 
If we fix the remaining $U(1)$ gauge degree of freedom in the Landau
gauge, the abelian gauge fields are given by 
\begin{eqnarray}
  \theta^{(i)}_\mu (s)= -\sum_{s'} D(s-s')
  \partial'_{\nu}f^{(i)}_{\nu \mu}(s').
\end{eqnarray} 
Extract the Dirac string from the field strength, satisfying
 $\sum_i k^{(i)}_\mu= \sum_i (1/2) \epsilon_{\mu\nu\rho\sigma}
  \partial_\nu n^{(i)}_{\rho\sigma} = 0$
\cite{polikarpov}:
\begin{eqnarray}
  f^{(i)}_{\mu\nu}(s) = 
  \bar{f}^{(i)}_{\mu\nu}(s)+ 2\pi n^{(i)}_{\mu\nu}(s),\ \ 
\left\{  \begin{array}{c}
   -\pi < \bar{f}^{(i)}_{\mu\nu}(s) \le \pi, \\
   n^{(i)}_{\mu\nu}(s) = 0,\pm 1,\pm 2, \pm 3.
  \end{array}\right.
\end{eqnarray}
The abelian gauge field from the  Dirac string is\cite{suzu95}
\begin{eqnarray}
  \theta^{Ds(i)}_\mu (s)= -2\pi \sum_{s'} D(s-s')
  \partial'_{\nu}n^{(i)}_{\nu \mu}(s') ,
\end{eqnarray} 
whereas the abelian gauge field from the photon is
\begin{eqnarray}
  \theta^{Ph(i)}_\mu (s)= - \sum_{s'} D(s-s')
  \partial'_{\nu}\bar{f}^{(i)}_{\nu \mu}(s') .
\end{eqnarray} 
Abelian link fields from the Dirac string and from the photon are
constructed as
\begin{eqnarray}
u^{Ds}(s,\mu)&=&\mbox{diag}(e^{i\theta^{Ds(1)}_\mu (s)}
                           ,e^{i\theta^{Ds(2)}_\mu (s)}
                           ,e^{i\theta^{Ds(3)}_\mu (s)}),   \\
u^{Ph}(s,\mu)&=&\mbox{diag}(e^{i\theta^{Ph(1)}_\mu (s)}
                           ,e^{i\theta^{Ph(2)}_\mu (s)}
                           ,e^{i\theta^{Ph(3)}_\mu (s)}).
\end{eqnarray}

\section{Simulations and results}
\label{results}

Now let us evaluate the inverse of the Wilson fermion matrix
\begin{eqnarray}
1-\kappa\sum_\mu[(1-\gamma_\mu)U(s,\mu)\delta_{s+\hat\mu,s'}
 + (1+\gamma_\mu)U^\dagger(s-\hat\mu,\mu)\delta_{s-\hat\mu,s'}]
\end{eqnarray}
for 
\begin{itemize}
  \item $U(s,\mu)$~~($SU(3)$),
  \item $u(s,\mu)$~~(abelian),
  \item $u^{Ds}(s,\mu)$~~(Dirac string),
  \item $u^{Ph}(s,\mu)$~~(photon),
  \item $C(s,\mu)u^{Ds}(s,\mu)$~~((matter) $\times$ (monopole))
  \item $C(s,\mu)u^{Ph}(s,\mu)$~~((matter) $\times$ (photon))
\end{itemize}
 and measure hadron mass spectra. $\kappa$ is the hopping parameter.
$C(s,\mu)u^{Ds}(s,\mu)$
($C(s,\mu)u^{Ph}(s,\mu)$) corresponds to a link variable in which the 
photon part (the monopole part) is excluded from the $SU(3)$ link variable.

\subsection{Simulations}
Our calculations were done on 4PE, 8PE and 16PE systems
of a vector parallel supercomputer Fujitsu VPP500.
The simulations were performed as follows:
\begin{itemize}
\item
     Lattice sizes and lattice spacings adopted are listed in Table
     \ref{conf_table}.
\item
     Antiperiodic (Periodic) boundary conditions in the time (space)
     direction.
\item
     Number of sweeps for thermalization is $3000 \sim 8000$.
     Number of sweeps to get independent configurations is  $1000 \sim 1200$.
     Number of configurations for average is listed in
     Table\ \ref{conf_table}.
\item
     The criterion of the MA gauge condition is 
     $\sum_s\mbox{$|$off diagonal part of $X(s)|$}< 10^{-7}$ where 
     $X$ is the operator to be diagonalized (Eq.(\ref{Xdiagonal})).
     An over-relaxation method is used.
\item
     The criterion of the solution of the Wilson matrix inversion is 
     $|$residue$|<10^{-5}$.
     The hopping parameter expansion and the CG method 
     with red-black preconditioning are used.
\item
     The mass fitting function is
     \begin{eqnarray}
      G(t) = c\cosh (m(t-N_t/2)).
       \label{propagator}
     \end{eqnarray}     
\end{itemize}

\begin{figure}
 \vspace{0.cm}
 \epsfxsize=0.7\textwidth
 \begin{center}
 \leavevmode
  \epsfbox{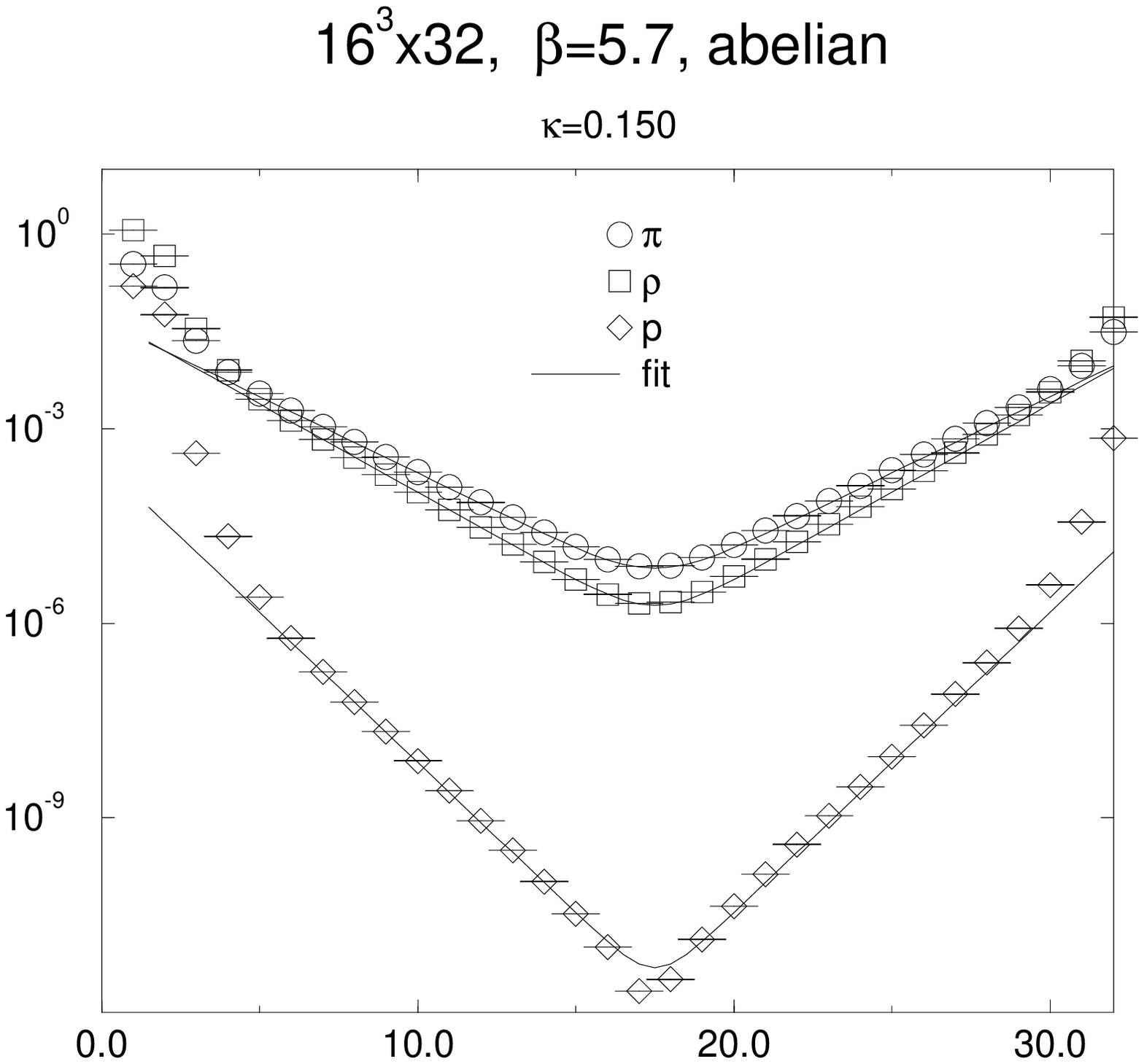}
 \end{center}
 \vspace{-1cm}
 \caption{
 Hadron propagators in the abelian gauge fields for $\kappa=0.150$
 at $\beta=5.7$.}
 \label{fig1}
 \end{figure}

\begin{figure}
 \vspace{0.cm}
 \epsfxsize=0.7\textwidth
 \begin{center}
 \leavevmode
  \epsfbox{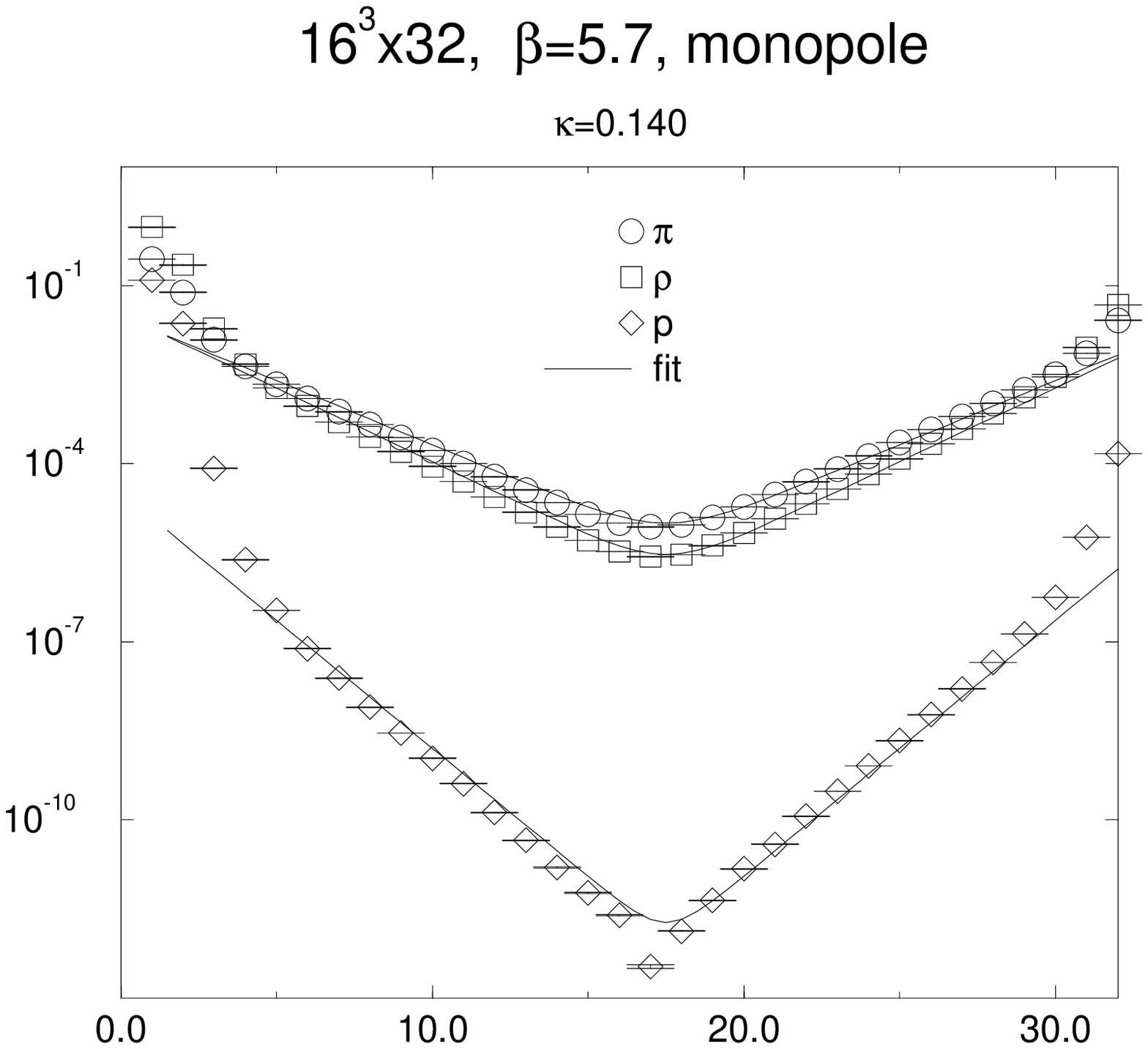}
 \end{center}
 \vspace{-1cm}
 \caption{
 Hadron propagators in the monopole gauge fields for $\kappa=0.140$
 at $\beta=5.7$.}
 \label{fig2}
 \end{figure}

Examples of hadron propagators in the abelian and the monopole
gauge fields are shown in Fig.\ref{fig1} and Fig.\ref{fig2}. Examples of
$\kappa$ adopted, the estimated masses of quark, $\pi$, $\rho$,
proton($p$) and $\Delta$ are listed in Table \ref{tab5.7}.

\subsection{The chiral limit}
Let us evaluate the chiral limit. 
It has been known that
the quenched approximation has a problem of 
chiral logs\cite{Gupta1}.
However, the linear relation between squared pion mass and quark mass 
holds for not very small quark mass\cite{Gupta2}.
We examine the relations in the abelian and in the monopole gauge fields
as well as in $SU(3)$ gauge field.

Usually the chiral limit is given by
demanding vanishing of the pion mass. It is derived by the following
fit:
\begin{eqnarray}
 m_\pi^2(\kappa) = A_1(1/2\kappa-1/2\kappa_c),
  \label{mpi2kappa}
\end{eqnarray}
where $\kappa_c$ is given by $m_\pi^2(\kappa_c) = 0$.
This method is valid when we can regard  $(1/2\kappa-1/2\kappa_c) \sim
m_q$.  Quenched chiral perturbative theory predicts 
\begin{eqnarray}
 m^{\overline{\rm MS}}_q = Z_m(1/2\kappa-1/2\kappa_c),
\end{eqnarray}
and it holds good in the usual lattice simulation with $SU(3)$ gauge
fields. 

\begin{figure}
\epsfxsize=0.7\textwidth
 \begin{center}
 \leavevmode
  \epsfbox{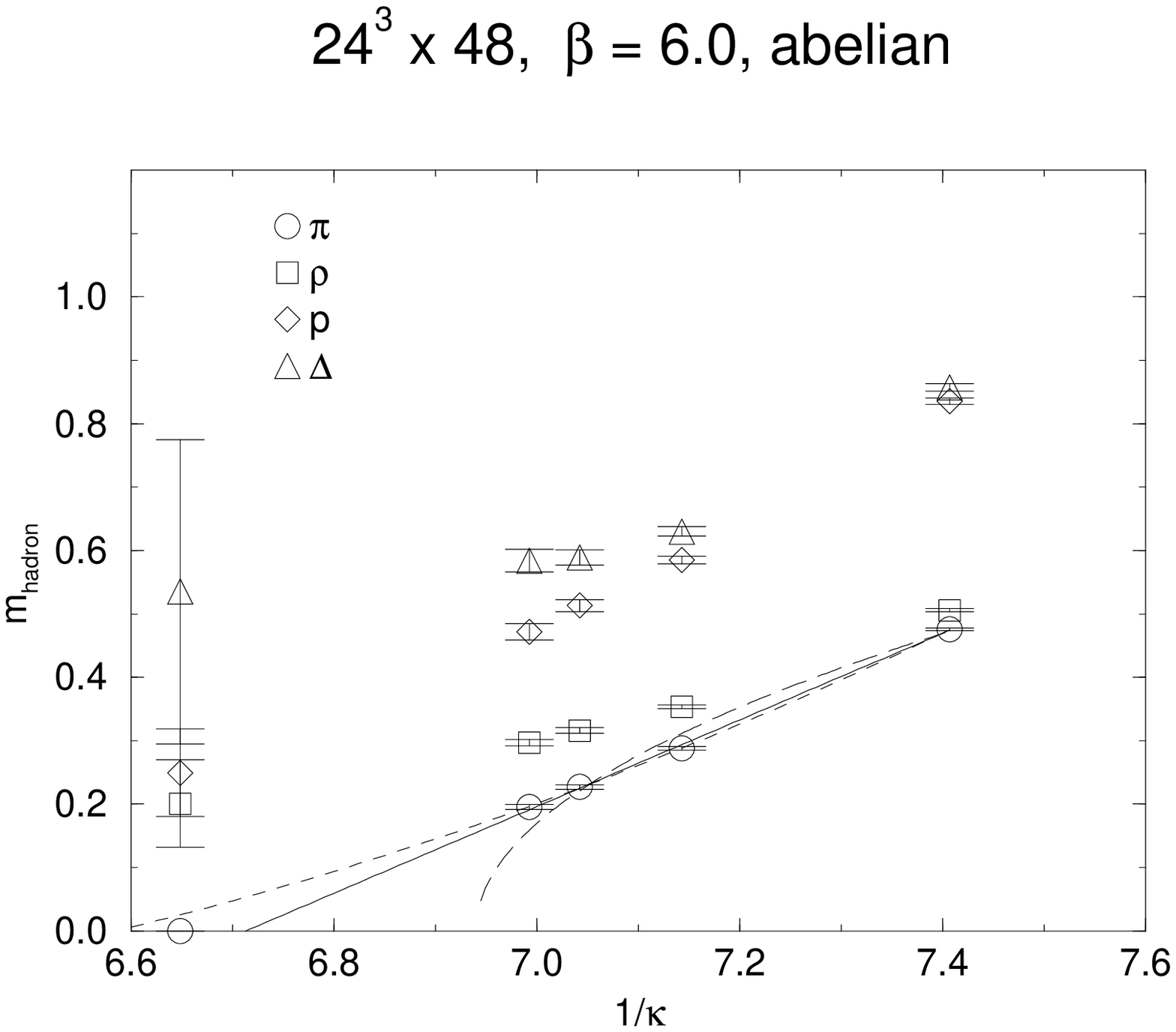}
 \end{center}
 \vspace{-1cm}
 \caption{
The masses of $\pi, \rho, p$,and $\Delta$ 
versus $1/\kappa$ on $24^3 \times 48 $ lattice at $\beta=6.0$
in the abelian gauge fields. }
\label{abelpv}
\end{figure}

\begin{figure}
\epsfxsize=0.7\textwidth
 \begin{center}
 \leavevmode
  \epsfbox{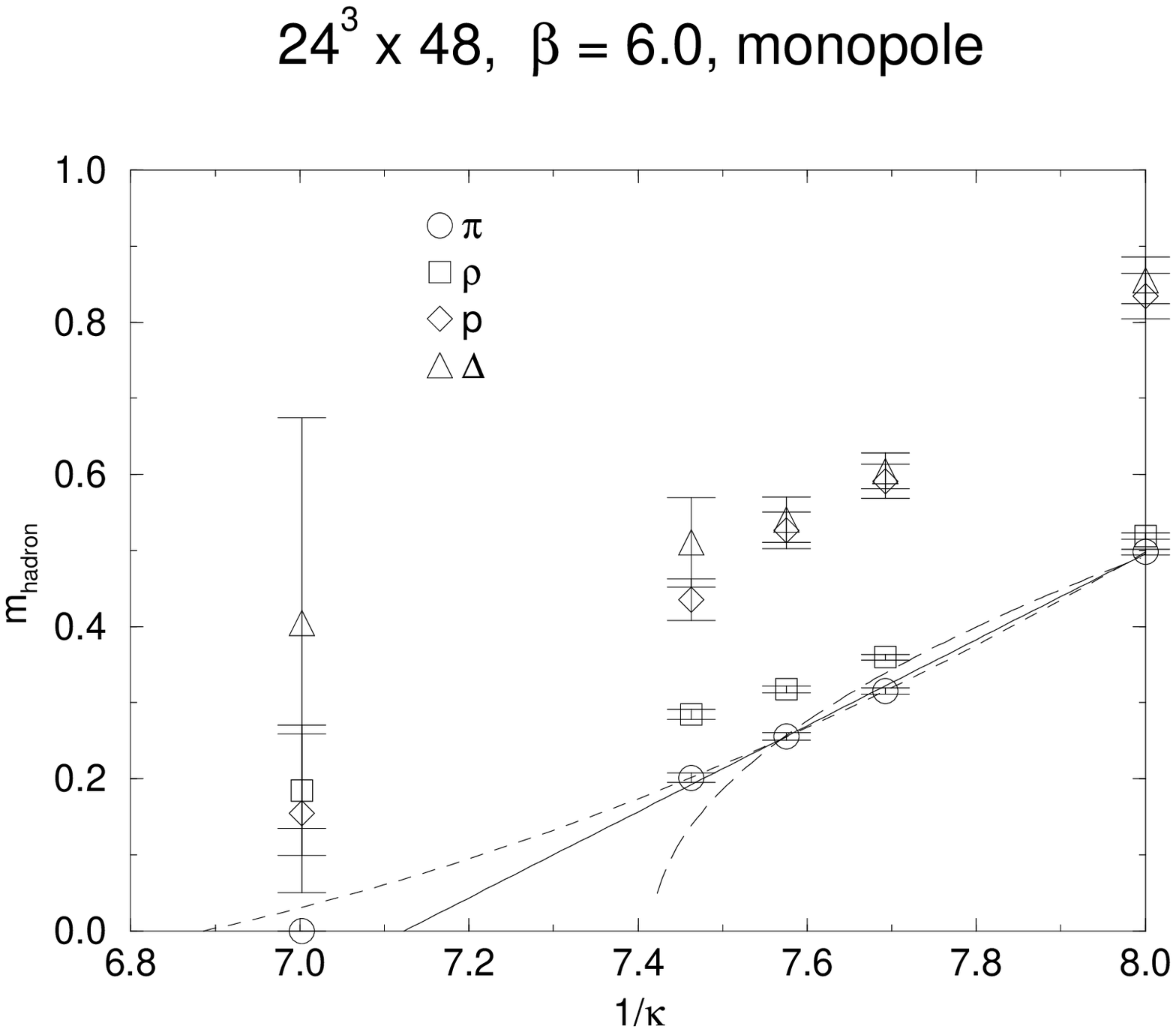}
 \end{center}
 \vspace{-1cm}
 \caption{
The masses of $\pi, \rho, p$,and $\Delta$ 
versus $1/\kappa$ on $24^3 \times 48 $ lattice at $\beta=6.0$
in the monopole gauge fields.
}
\label{monopv}
\end{figure}

Howeve in the abelian and the monopole gauge fields, the $\kappa$
dependences of  hadron and quark masses are not known.
First, we use three types of fits for the
pion mass; i.e., the fit (\ref{mpi2kappa}) and
\begin{eqnarray}
  m(\kappa) &=& B_0 + B_1(1/\kappa),
  \label{fit1}  \\
  m(\kappa) &=& C_0 + C_1(1/\kappa) + C_2(1/\kappa)^2.
  \label{fit2}
\end{eqnarray}
The latter two types of the fits are used also 
for $\rho, p$ and $\Delta$.
 The critical hopping parameter $\kappa_c$ is obtained by the condition 
$m_\pi(\kappa_c)=0$. The masses of $\pi, \rho, p$ and $\Delta$ are
plotted versus $1/\kappa$ in Fig.\ \ref{abelpv} for the abelian
background and Fig.\ \ref{monopv} for the monopole one. We find a
strange dependence of the hadron masses on $\kappa$.
In the figures, a long dashed line, a solid line and a dashed line 
denote the fits (\ref{mpi2kappa}), (\ref{fit1}) and
(\ref{fit2}), respectively.
For reference, we have plotted the masses at the chiral limit
given by the average of the fits (\ref{fit1}) and (\ref{fit2}) at the
most left side ($\kappa_c$ is also given by the average) of  Fig.\
\ref{abelpv} and Fig.\ \ref{monopv}. 
The error of the hadron masses at the chiral limit is
estimated by the summation of the systematic error 
from the indefiniteness of the fitting functions
and the statistical error given by the $\chi^2$ fit.
Unfortunately, these fittings have large ambiguities with respect to
the masses at the chiral limit.  The  $\chi^2/d.o.f.$ are very large.
We need, hence, another fit to extrapolate the
hadron masses toward the chiral limit without using $\kappa$ dependence.

We can define the quark mass from the axial Ward identity as
follows\cite{gupta}:
\begin{eqnarray}
 \frac{Z_P}{Z_A}m_q = -\lim_{t \rightarrow \infty}
  m_\pi \frac{\langle 0|A_4(t)P(0)|0\rangle }{\langle 0|P(t)P(0)|0\rangle }
  \label{mq^WI}
\end{eqnarray}
where $A_4(t)$ is the 4-th component of an axial vector current;
$A_\mu(t) = \bar{\Psi}(t)\gamma_\mu\gamma_5\Psi(0)$ and $P(t)$ is a 
pseudo scalar current;
$P(t) = \bar{\Psi}(t)\gamma_5\Psi(0)$.
Generally the renormalization constants for the axial vector current
$Z_A$  and for the pseudo scalar current $Z_P$  are not unity for
finite lattice spacing\cite{bochi85}.
However we approximated $Z_P/Z_A$ by unity because
its value including non-perturbative effects is not known.

\begin{figure}
\epsfxsize=0.75\textwidth
 \begin{center}
 \leavevmode
  \epsfbox{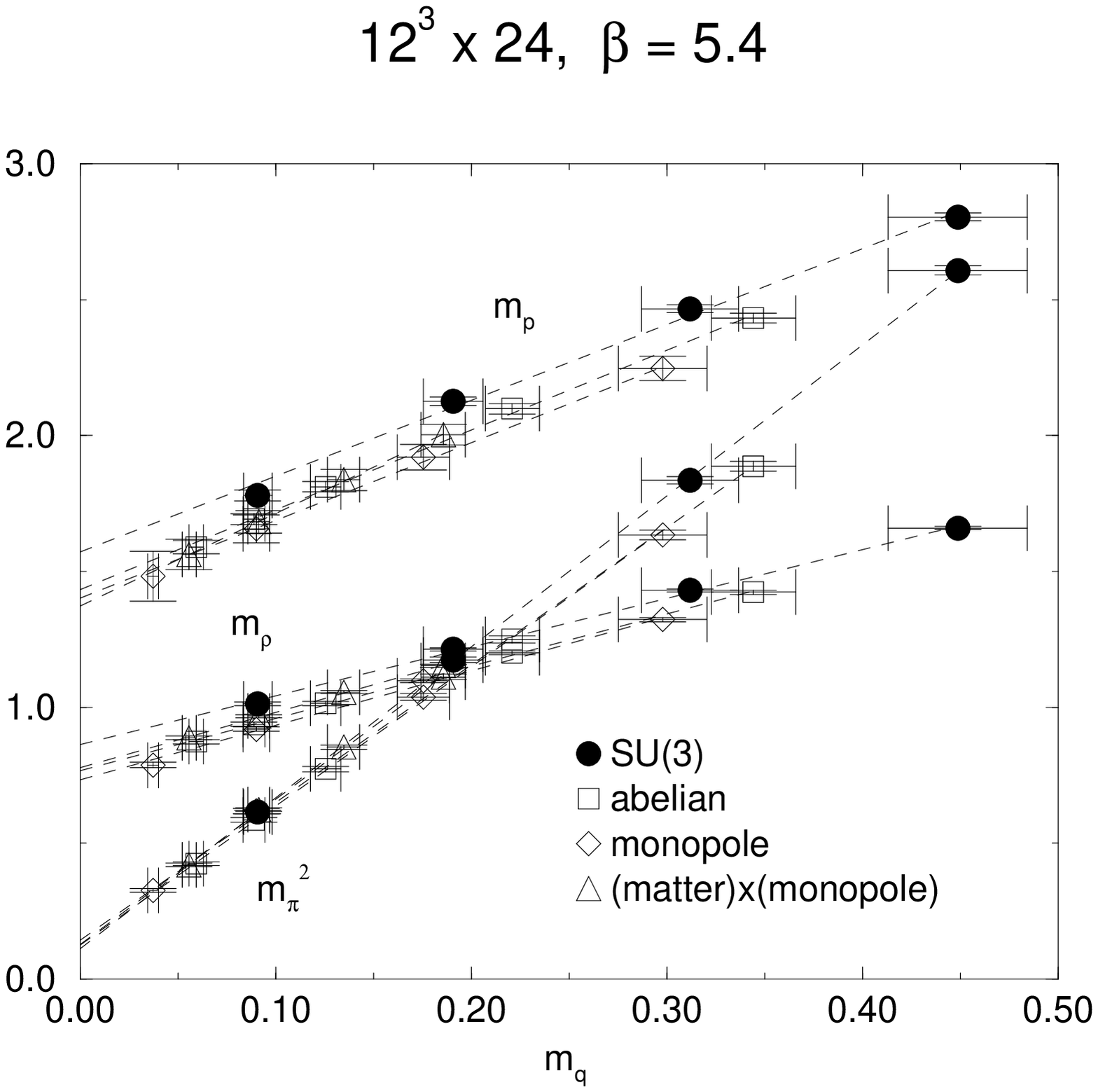}
 \end{center}
\vspace{-1cm}

 \caption{
$m_\pi^2, \rho$ and $p$ versus $m_q$.
}
\label{S54pvp_mq_p}
\end{figure}

\begin{figure}
\epsfxsize=0.75\textwidth
 \begin{center}
 \leavevmode
  \epsfbox{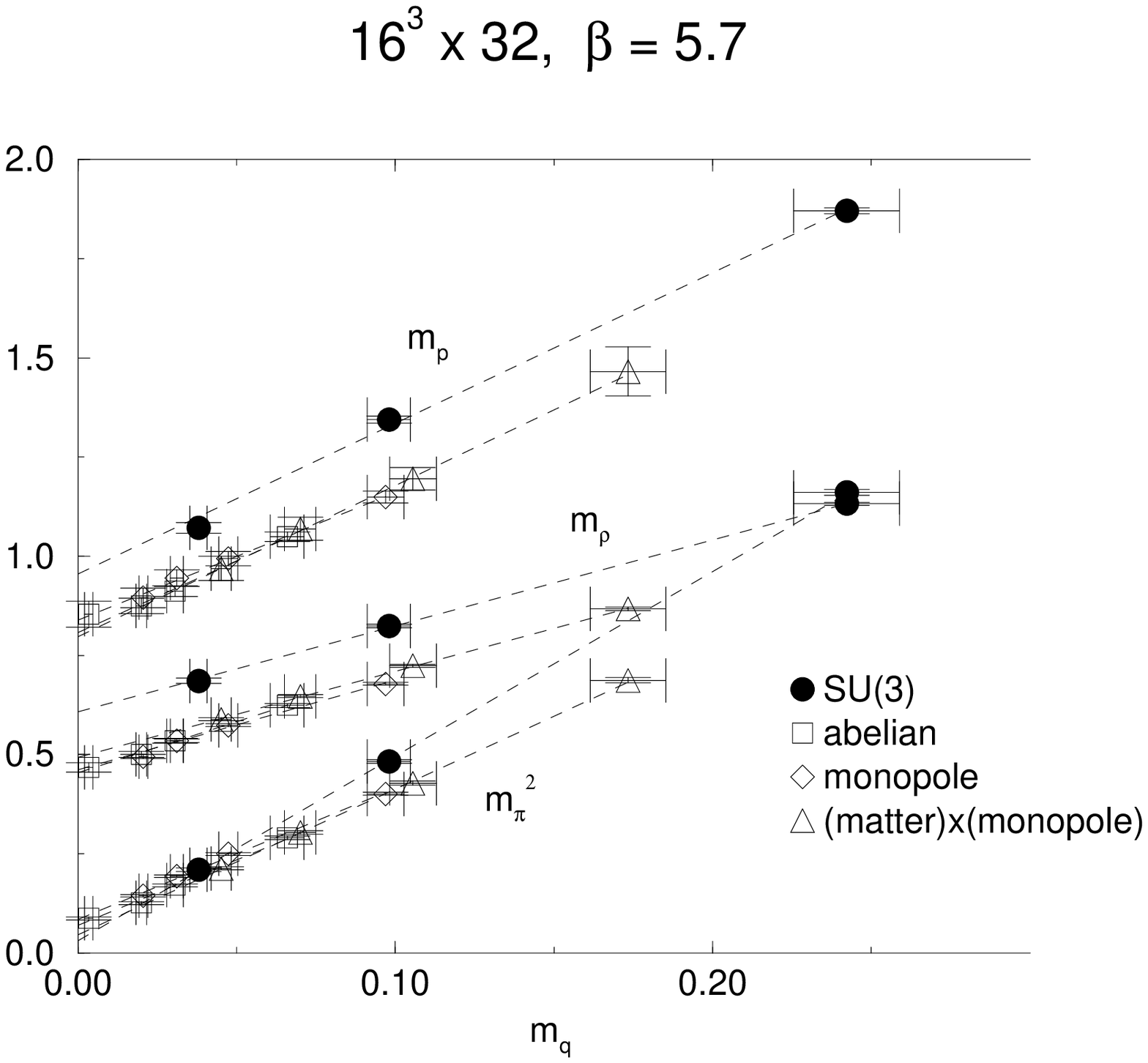}
 \end{center}
\vspace{-1cm}

 \caption{
$m_\pi^2, \rho$ and $p$ versus $m_q$.
}
\label{M57pvp_mq_p}
\end{figure}

Fig.\ \ref{S54pvp_mq_p} and Fig.\ \ref{M57pvp_mq_p} show the hadron
masses versus the quark mass defined 
by Eq.(\ref{mq^WI}). We observe the linear behaviors 
\begin{eqnarray}
 m^2_\pi &=& A_1 m_q  \label{mpi2mq} + A_0,  \label{m2pimq} \\
 m_{\rho, p, \Delta} &=& B_{\rho, p, \Delta} m_q
  + m^c_{\rho, p, \Delta} \label{mHmq}
\end{eqnarray}
for $SU(3)$, abelian, monopole and (matter)$\times$(monopole)
\footnote{
The r.h.s. of Eq.(\ref{mq^WI}) is almost
proportional to $m_\pi^2$ because of Lorentz covariance
\cite{iwasaki86}.}.
$m^c$ is the mass at the chiral limit defined by zero quark mass.
The masses at the chiral limit
after the abelian projection 
are slightly smaller than those of $SU(3)$.

\subsection{Gross features of hadron masses}
\label{gross}
\begin{figure}
\epsfxsize=0.75\textwidth
 \begin{center}
 \leavevmode
  \epsfbox{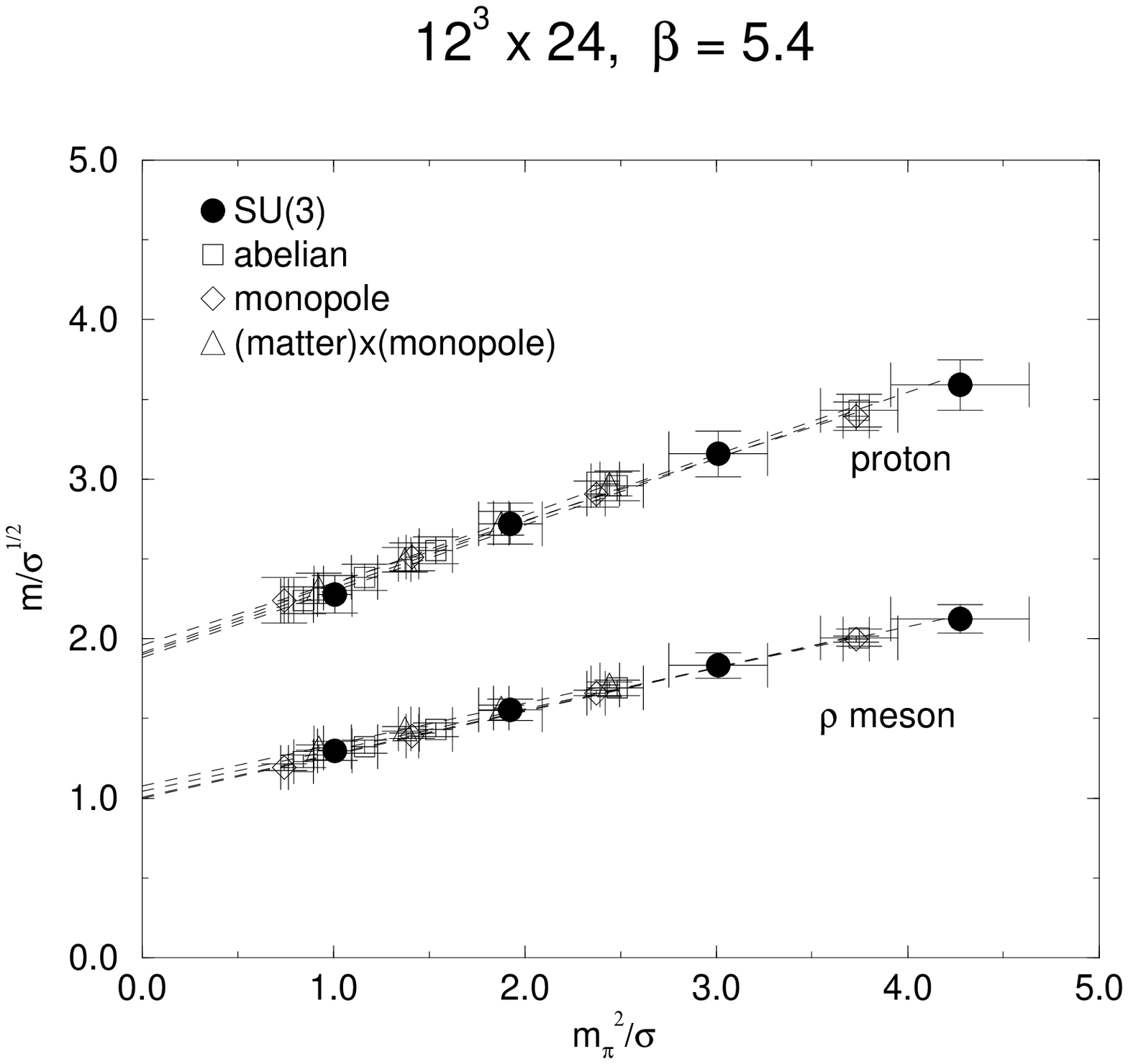}
 \end{center}
\vspace{-1cm}
 \caption{
The $m_\rho/\sigma^{1/2}$ and $m_p/\sigma^{1/2}$ versus
 $m_\pi^2/\sigma$.
}
\label{S54ms_p2vp1_p}
\end{figure}

\begin{figure}
\epsfxsize=0.75\textwidth
 \begin{center}
 \leavevmode
  \epsfbox{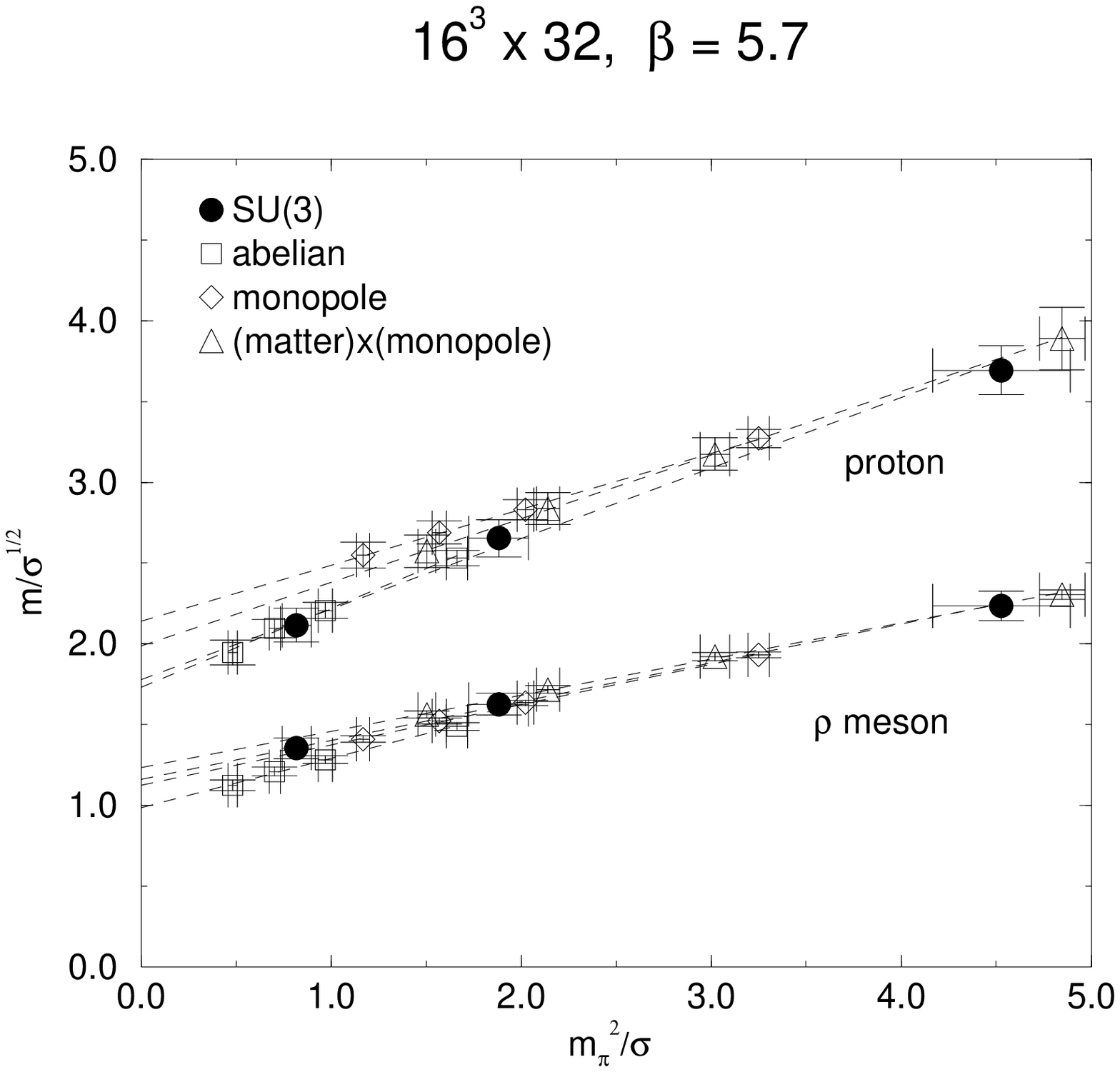}
 \end{center}
\vspace{-1cm}
 \caption{
The $m_\rho/\sigma^{1/2}$ and $m_p/\sigma^{1/2}$ versus
 $m_\pi^2/\sigma$.
}
\label{M57ms_p2vp1_p}
\end{figure}

Fig.\ \ref{S54ms_p2vp1_p} and Fig.\ \ref{M57ms_p2vp1_p} 
 show dimensionless ratios
$m_\rho/\sqrt{\sigma}$ and $m_p/\sqrt{\sigma}$ versus
$m_\pi^2/\sigma$ where $\sigma$ is the string
tension in each gauge field. We can see that all ratios of the three
types, in the abelian, the monopole and the (matter)$\times$(monopole)
gauge fields, are equal or very close to the values of $SU(3)$.
In Fig.\ \ref{S54ms_p2vp1_p} and Fig.\ \ref{M57ms_p2vp1_p}, long
dashed lines represent the linear fits with $m_\pi^2$:
\begin{eqnarray}
 m_{\rho,p,\Delta}/\sqrt{\sigma} = C_{\rho, p, \Delta} m_\pi^2/\sigma
  + \bar{m}^c_{\rho, p, \Delta}/\sqrt{\sigma}.
  \label{mHmpi2}
\end{eqnarray}

Now we use the fit (\ref{mHmq}) or the fit (\ref{mHmpi2})
for extrapolation toward the chiral limit.
The masses normalized by the squared root of the string tension 
and $\chi^2/d.o.f$ of the fits are listed in
Table \ref{chiral_mqWI} and Table \ref{ms_p2vpd}. 
The masses of $\rho$,
$p$ and $\Delta$ are given by setting $m_q = 0$ for the fit (\ref{mHmq})
or $m_\pi^2=0$ for the fit (\ref{mHmpi2}).
The two fits give almost the same results, $m^c_{\rho, p, \Delta}$ and 
$\bar{m}^c_{\rho, p, \Delta}$. 
This is consistent with smallness of $A_0$ in (\ref{m2pimq}) as shown in 
Table \ref{chiral_mqWI}. Hence we call them as hadron masses in the
following. 
We observe {\it abelian dominance} and {\it monopole
dominance} for the hadron masses normalized by the squared root of the
string tension 
\begin{eqnarray}
 \frac{m_\rho^{SU(3)}} {\sqrt{\sigma^{SU(3)}}} \simeq
  \frac{m_\rho^{\rm abel}} {\sqrt{\sigma^{\rm abel}}}
  \simeq  \frac{m_\rho^{\rm mono}} {\sqrt{\sigma^{\rm mono}}}
  \simeq  \frac{m_\rho^{{\rm (matt)} \times {\rm (mono)}}}
  {\sqrt{\sigma^{{\rm (matt)} \times {\rm(mono)}}}} \\
 \frac{m_B^{SU(3)}} {\sqrt{\sigma^{SU(3)}}} \simeq
  \frac{m_B^{\rm abel}} {\sqrt{\sigma^{\rm abel}}}
  \lsim  \frac{m_B^{{\rm (matt)} \times {\rm (mono)}}}
  {\sqrt{\sigma^{{\rm (matt)} \times {\rm(mono)}}}}
  \lsim  \frac{m_B^{\rm mono}} {\sqrt{\sigma^{\rm mono}}}
\end{eqnarray}
where $B$ denotes $p$ and $\Delta$.

The ratios $m_p/\sqrt{\sigma}$ and $m_\Delta/\sqrt{\sigma}$
in the (matter)$\times$(monopole) and monopole gauge fields are
slightly larger than those of $SU(3)$. 
Comparing Fig.7 and Fig.8, the tendency is clearer in the data on
$16^3\times32$ lattice  
at $\beta=5.7$. Thus, the deviation may be due to the difference of the
short range structure  of the gauge fields. 
We have also found similar behaviors by measuring
$\rho$ or the baryon masses at different $\beta$s 
on the same lattice size.
As $\beta$ becomes smaller, 
the abelian and monopole dominances become more conspicuous.

\begin{figure}
\epsfxsize=0.75\textwidth
 \begin{center}
 \leavevmode
  \epsfbox{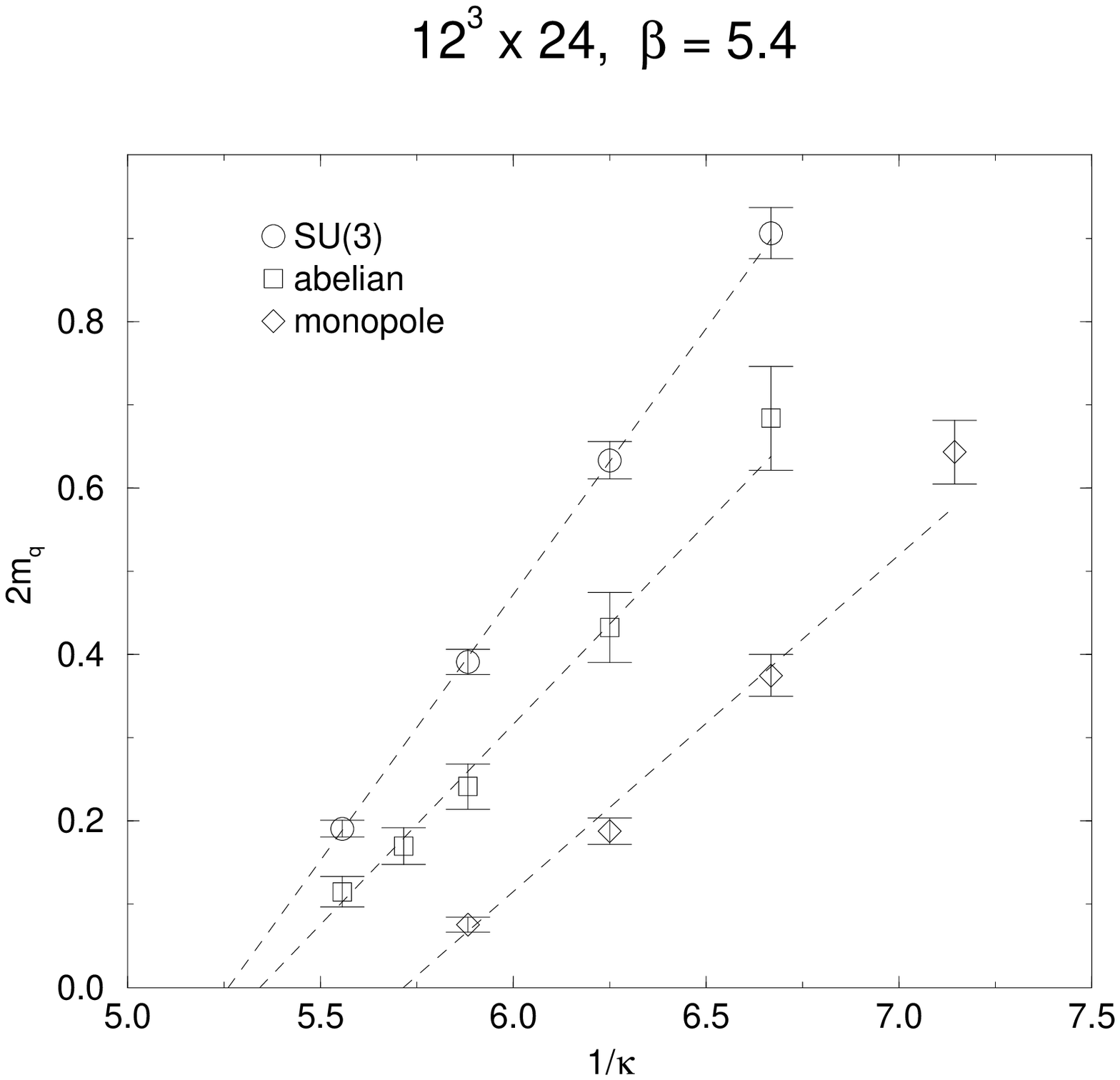}
 \end{center}
\vspace{-1cm}

 \caption{
$2m_q$ versus $1/\kappa.$
}
\label{S54mq_k_p}
\end{figure}

We have plotted the quark mass versus $1/\kappa$ in Fig.\
\ref{S54mq_k_p}.
We see that  $(1/2\kappa - 1/2\kappa_c)$ is not perfectly proportional
to the quark mass derived by Eq.(\ref{mq^WI}) in the abelian and the
monopole gauge fields.
We need some higher order terms such as $O((1/\kappa)^2)$.
This shows that 
$(1/2\kappa-1/2\kappa_c)$ cannot be recognized 
straightforwardly as the quark mass in the abelian and the monopole
gauge fields.
However 
in the case of $SU(3)$, the fits using (\ref{fit1}),
(\ref{mHmq}) and (\ref{mHmpi2})  give almost the same value of the
masses for $\rho, p$ and $\Delta$ (Table \ref{chiral_mqWI} and
Table \ref{ms_p2vpd}).

\subsection{Hadron spectra in gauge fields without monopoles }
Next, we investigate hadron spectra in the absence of monopoles.   
Hadronic correlators are measured 
in the monopoleless abelian field, i.e. in the photon part $u^{Ph}$
and also in the monopoleless $SU(3)$ gauge fields
$Cu^{Ph}$.  The latter 
is (matter)$\times$(abelian part)/(monopole part) = 
(matter)$\times$(photon part).

\begin{figure}
\epsfxsize=0.7\textwidth
 \begin{center}
 \leavevmode
  \epsfbox{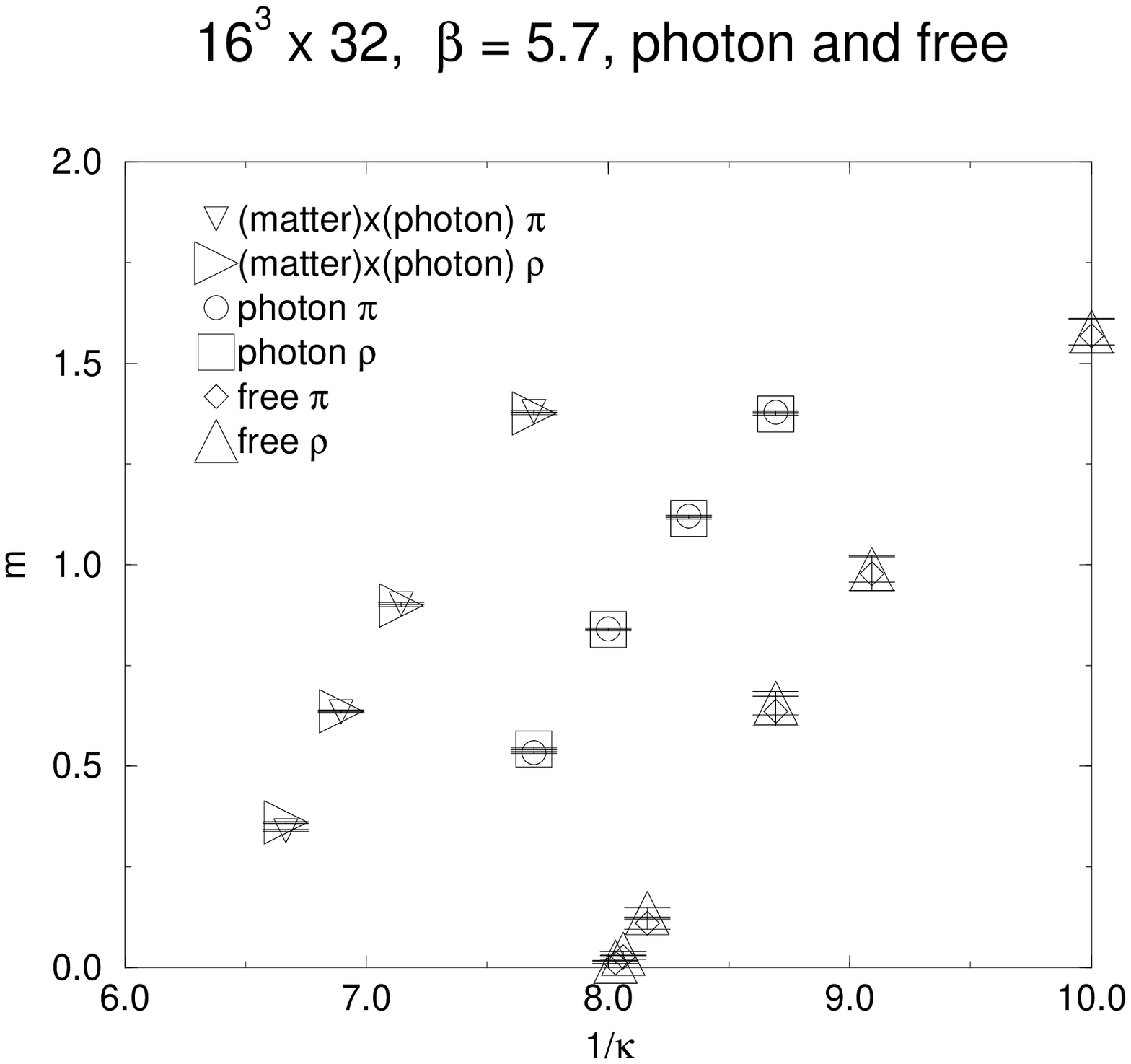}
\end{center}
 \vspace{-1cm}
 \caption{
The squared masses of $\pi$ and $\rho$ versus $1/\kappa$ in the
 free, the photon and the (matter)$\times$(photon) gauge fields.
}
\label{photon_free}
\end{figure}

Fig.\ \ref{photon_free} shows the masses of $\pi$ and 
$\rho$ versus $1/\kappa$ in the photon, the (matter)$\times$(photon)
and the free fields.
The estimated masses of $\pi$, $\rho$, $p$ and $\Delta$ in each
field are listed in
Table\ \ref{Tbl.ph-fr}. 

As seen from Fig.\ \ref{photon_free} and Table\ \ref{Tbl.ph-fr},
the photon contribution does not produce any dynamical mass generation 
($\kappa_c= 0.140(2)$ for $16^3 \times 32$ lattice, $\beta = 5.7$),
namely, 
\begin{eqnarray}
 m_\pi \simeq m_\rho, &\ \ &  m_p \simeq m_\Delta.
\end{eqnarray}
The mass ratio is given by
\begin{eqnarray}
  m_{p}/m_{\rho} \sim  m_{\Delta}/m_{\rho} \sim 3/2
\end{eqnarray}
in the case of the photon. 
The hadron masses are simply in proportion to the number of the bare
quarks. The behavior is similar to those in
the free case, although $\kappa_c= 0.125$.

Similar behaviors are seen in the monopoleless $SU(3)$ gauge fields.
It does not either
generate the mass gap  ($\kappa_c = 0.157(2)$) as in the previous photon
case.  We find that
{\it the dynamical mass generation cannot be produced without
monopoles.} 
The matter field is not crucial for the dynamical mass generation.

\section{Summary and Remarks}
\label{summary}
The hadron spectra in abelian, monopole, photon,
(matter)$\times$(monopole) and (matter)$\times$(photon) fields
are  studied. 
The following results are obtained:
\begin{itemize}
 \item
      The PCAC relation holds good in the abelian, the monopole and the
      (matter) $\times$ (monopole) fields. The squared pion mass
      is well proportional to the quark mass derived from the axial Ward 
      identity. 
      The perturbative or empirical relation
      $m_q \sim 1/2\kappa-1/2\kappa_c$, does not hold in the abelian and 
      the monopole gauge fields. 

 \item 
      The ratios of the hadron mass to the square root of the
      string tension of 
      the abelian, the monopole and the (matter) $\times$ (monopole) 
      are similar to those of $SU(3)$.  
      In this sense, 
      we have found the abelian dominance and the monopole dominance for
      the hadron spectra. 
      However, some indications of the difference which may be
      interpreted as the shorter range 
      effects are also found.

 \item 
      No sizable dynamical mass generation is seen in the 
      measurements both in the abelian and in $SU(3)$ gauge fields
      without monopoles. These results suggest that monopoles are
      crucial for the dynamical mass generation and that the photon part 
      and the matter part are not important.
       
\end{itemize}

The data are consistent with the results of simulations in quenched 
$SU(2)$ QCD with Kogut-Susskind fermions\cite{miya95}.

We would like to thank Atsushi Nakamura of Hiroshima University for
his kind help in the solver coding. We thank Dr. Shoji Hashimoto of 
High Energy Accelerator Research Organization (KEK)
for his kind help in coding using FFT.
This work is supported by the Supurcomputer Project (No.97-17) of KEK.
The calculations are also performed on the Fujitsu VPP/500 at the
institute of Physical and Chemical Research (RIKEN). 
This work is financially supported by JSPS \\
for Exploratory Research (No.09874060) (T.S.)
and also supported by the Grant in Aid
for Scientific Research C (2)(No 0760411) (O.M.)
by the Ministry of Education, Science and Culture.

\input epsf

\begin{table}[tbh]
\begin{center}
\vspace{0.5cm}
$\begin{array}{|cc|ccc|ccc|ccc|} \hline
&&\multicolumn{3}{|c|}{SU(3):} &
       \multicolumn{3}{|c|}{$abelian:$} &
            \multicolumn{3}{|c|}{$monopole:$}        \\ \hline
L_s^3 \times L_t & \beta & a$[fm]$ & aL_s$[fm]$ & \#
                          & a$[fm]$ & aL_s$[fm]$ & \#
                           & a$[fm]$ & aL_s$[fm]$ & \#  \\ \hline
24^3 \times 48 &  6.0   &          &          &  
                        & .101(1)  & 2.42(2)  &  52
                        & .083(1)  & 1.98(2)  &  52   \\
16^3 \times 32 &  5.7   & .227(8)  & 3.63(13) &  32
                        & .189(2)  & 3.02(4)  & 150
                        & .158(1)  & 2.52(1)  & 100   \\
16^3 \times 32 &  5.8   & .207(9)  & 3.31(14) &  48
                        & .154(3)  & 2.47(4)  &  78
                        & .120(1)  & 1.91(1)  &  48   \\
12^3 \times 24 &  5.4   & .350(13) & 4.20(16) &  60
                        & .318(7)  & 3.82(8)  & 110
                        & .297(16) & 3.56(19) & 110   \\
12^3 \times 24 &  5.53  &          &          &  
                        & .271(1)  & 3.25(2)  &  70
                        &          &          &       \\
12^3 \times 24 &  5.6   &          &          &
                        & .235(2)  & 2.82(2)  & 100
                        & .205(1)  & 2.46(1)  & 110   \\
12^3 \times 24 &  5.66  & .254(5)  & 3.05(6)  &  48
                        & .210(3)  & 2.52(3)  & 102
                        & .175(1)  & 2.10(1)  & 102   \\ \hline
\end{array}$
\\
\vspace{0.5cm}
\caption{Physical lattice size (from the $SU(3)$, abelian and monopole
 string tension for $\sigma^{1/2}=$440MeV) 
 and the number of configurations on each of the lattice.}
\label{conf_table}
\end{center}
\end{table}
\newpage
\begin{table}[htb]
\vspace{-3.5cm}
\begin{center} 
$\begin{array}{|l|c|c|c|c|c|} 
 \hline
\multicolumn{6}{|l|}{SU(3):} \\ \hline 
\kappa  &   m_q       &   m_{\pi}   & m_{\rho}  &  m_p       & m_\Delta \\\hline
0.150   &  .2422(166) &   1.069(04) & 1.118(05) & 1.870(08)  &  1.911(09)\\
0.155   &  .0980(067) &   0.888(03) & 0.956(12) & 1.344(09)  &  1.436(10)\\
0.160   &  .0380(026) &   0.692(07) & 0.814(10) & 1.072(13)  &  1.215(18)\\
        &  .000       &   0.180(07) & 0.605(08) & 0.936(27)  &  1.092(18)\\ \hline
	  \hline
\multicolumn{6}{|l|}{$abelian:$} \\ \hline 
\kappa & m_q         & m_{\pi}    & m_{\rho}  & m_p       & m_\Delta  \\ \hline
0.150  &  .0660(53)  & 0.543(03)  &  .624(05) & 1.049(11) &  1.111(13) \\
0.155  &  .0307(25)  & 0.415(04)  &  .535(06) &  .912(14) &  1.001(18) \\
0.1575 &  .0200(17)  & 0.354(03)  &  .501(08) &  .874(19) &   .959(29) \\
0.160  &  .0035(13)  & 0.293(04)  &  .468(12) &  .853(33) &   .892(52) \\
       &  .000       & 0.267(05)  &  .456(08) &  .821(24) &   .888(43) \\ \hline
	  \hline
\multicolumn{6}{|l|}{$monopole:$} \\ \hline 
\kappa  & m_q       & m_{\pi}  & m_{\rho} & m_p       & m_\Delta  \\\hline
0.135   &  .0970(57)& .633(03) & .678(32) & 1.14(15)  & 1.18(17)  \\
0.140   &  .0473(29)& .499(03) & .574(34) &  .99(17)  & 1.04(21)  \\
0.1425  &  .0311(20)& .440(04) & .535(39) &  .94(21)  & 1.00(26)  \\
0.145   &  .0216(13)& .379(04) & .495(50) &  .89(24)  &  .98(35)  \\
        &  .000     & .281(11) & .452(10) &  .83(17)  &  .92(19)  \\ \hline
\end{array}$
\end{center}
\caption{
The quark mass and hadron masses for $16^3 \times 32$ at $\beta=5.7$.
}
\label{tab5.7}
\end{table}
\begin{table}

 \begin{center}
 $\begin{array}{|c|ccccc|}\hline
\multicolumn{6}{c}{$fit using $m_q^{WI}$ for $12^3 \times 24$ at $\beta=5.4} \\ \hline
     & $SU(3)$ & SU(3)(1/\kappa)
      & $abelian$ & $monopole$ & $(matter)$\times$(monopole)$ \\   \hline
A_0                   & .110(1) & ---     & .123(2)  & .139(5)  &  .123(4) \\
\chi^2/d.o.f.         &  .001   & 1.819   & .020     &    .097  &    .042  \\  \hline
m_\rho/\sqrt{\sigma}  & 1.09(06)& 1.02(08)& 1.07(03) & 1.09(03) &1.15(01)  \\
\chi^2/d.o.f.         & .081    &   .990  & .121     &    .397  &     .065 \\ \hline
m_p/\sqrt{\sigma}     & 1.97(12)& 1.83(18)& 2.00(07) & 2.09(05) &2.03(02)  \\
\chi^2/d.o.f.         & .278    &   1.950 & .202     &   .167   &     .010 \\ \hline
m_\Delta/\sqrt{\sigma}& 2.17(11)& 2.07(14)& 2.12(08) & 2.45(04) & 2.07(07) \\
\chi^2/d.o.f.         & .049    &    .210 & .342     &   .027   &     .710 \\ \hline
 \end{array}$
\end{center}

\begin{center}
$\begin{array}{|c|ccccc|}\hline
\multicolumn{6}{c}{$fit using $m_q^{WI}$ for $16^3 \times 32$ at $\beta=5.7} \\ \hline
  & SU(3) & SU(3)(1/\kappa)
   & $abelian$ & $monopole$ & $(matter)$\times$(monopole)$ \\   \hline
A_0                   &  .032(2)& ---     &  .071(3)&  .079(6)&  .047(2) \\
\chi^2/d.o.f.         &  .010   & 2.242   &  1.078  &  .790   &  .015    \\  \hline
m_\rho/\sqrt{\sigma}  & 1.20(06)& 1.15(08)& 1.08(03)& 1.29(03)& 1.31(02) \\
\chi^2/d.o.f.         &  .041   &  .022   &  .661   & 1.475   & .023     \\ \hline
m_p/\sqrt{\sigma}     & 1.85(12)& 1.74(18)& 1.95(08)& 2.37(06)& 2.12(04) \\
\chi^2/d.o.f.         &  .377   &  1.04   &  5.824  &  .742   & .025     \\ \hline
m_\Delta/\sqrt{\sigma}& 2.16(12)& 2.08(14)& 2.11(13)& 2.63(06)& 2.31(05) \\
\chi^2/d.o.f.         &  .106   &  .020   &  .373   & .331    & .037     \\ \hline
\end{array}$
\end{center}

\caption{
The pion intercept and 
the ratios between the masses at the chiral limit and 
the squared root of the string tension.
$SU(3)(1/\kappa)$ denotes the masses that given by fits as
linear function of $1/\kappa$ .
}
\label{chiral_mqWI}
\end{table}
\begin{table}

 \begin{center}
 $\begin{array}{|c|ccccc|}\hline
\multicolumn{6}{c}{$fit using $m_\pi^2/\sqrt{\sigma}$\ 
 for $12^3 \times 24$ at $\beta=5.4}\\ \hline
   & SU(3) & SU(3)(1/\kappa)
                 & $abelian$ & $monopole$ & $(matter)$\times$(monopole)$  \\ \hline
m_\rho/\sqrt{\sigma}  & 1.04(12) & 1.02(08)& 1.01(06) & 1.00(04) & 1.08(04) \\
\chi^2/d.o.f.       &     .043 &     .99 &     .051 &     .651 &     .010 \\ \hline
m_p/\sqrt{\sigma}     & 1.88(23) & 1.83(18)& 1.90(12) & 1.96(14) & 1.91(16) \\
\chi^2/d.o.f.       &     .137 &    1.95 &     .081 &     .353 &     .108 \\ \hline
m_\Delta/\sqrt{\sigma}& 2.10(21) & 2.07(14)& 2.02(13) & 2.36(17) & 1.94(18) \\
\chi^2/d.o.f.       &    .024  &    .21  &     .117 &     .382 &     1.084\\ \hline
 \end{array}$
 \end{center}

\begin{center}
$\begin{array}{|c|ccccc|}\hline
\multicolumn{6}{c}{$fit using $m_\pi^2/\sqrt{\sigma}$\ 
 for $16^3 \times 32$ at $\beta=5.7}\\ \hline
  & SU(3) & SU(3)(1/\kappa)
   & $abelian$ & $monopole$ & $(matter)$\times$(monopole)$ \\
   \hline
m_\rho/\sqrt{\sigma}  & 1.16(10)& 1.15(08) & .99(05) & 1.13(04)& 1.23(05)  \\
\chi^2/d.o.f.       &  .037   &  .022    &  .248   &  .208   &  .214     \\  \hline
m_p/\sqrt{\sigma}     & 1.78(19)& 1.74(18) & 1.73(10)& 2.14(12)& 1.99(19)  \\
\chi^2/d.o.f.       &  .216   &  1.04    &  .520   & .099    & .028      \\  \hline
m_\Delta/\sqrt{\sigma}& 2.10(19)& 2.08(14) & 2.01(14)& 2.43(15)& 2.19(19)  \\
\chi^2/d.o.f.       &  .068   &  .020    & 8.413   & 2.993   &  .114     \\  \hline
\end{array}$
\end{center}

\caption{
The ratios between the masses at the chiral limit and 
the squared root of the string tension.
$SU(3)(1/\kappa)$ denotes the masses that given by fits as
linear function of $1/\kappa$ .
}
\label{ms_p2vpd}
\end{table}
\begin{table}[htb]
\vspace{-.5cm}
\begin{center} 
$\begin{array}{|l|c|c|c|c|} 
\hline
\multicolumn{5}{|l|}{$photon:$} \\ \hline 
\kappa          & m_{\pi} & m_{\rho} & m_p & m_\Delta \\\hline
 .1150 &1.379(2)&1.374(2)& 2.182(4)& 2.181(4) \\
 .1200 &1.120(2)&1.115(2)& 1.790(4)& 1.789(4) \\
 .1250 & .841(2)& .839(2)& 1.355(4)& 1.355(4) \\
 .1300 & .534(2)& .543(2)&  .872(4)&  .875(4) \\
 .140(2)(\kappa_c) & 0.0 &  .023(10) & 0.033(16) & 0.042(12) \\ \hline \hline
\multicolumn{5}{|l|}{$(matter)$\times$(photon)$} \\ \hline 
\kappa & m_{\pi}   & m_{\rho}   & m_p       & m_\Delta  \\ \hline
0.130 & 1.380(2) &  1.375(3) &  2.192(6) & 2.193(6) \\
0.140 &  .903(2) &   .898(2) &  1.451(5) & 1.450(5) \\
0.145 &  .634(2) &   .636(2) &  1.023(4) & 1.024(5) \\
0.150 &  .340(1) &   .360(2) &   .578(4) &  .588(4) \\ \hline \hline
\multicolumn{5}{|l|}{$free:$} \\ \hline 
\kappa & m_{\pi}   & m_{\rho}   & m_p       & m_\Delta  \\ \hline
0.10   & 1.576(21) &  1.581(18) & 2.500(42) & 2.502(41) \\
0.11   &  .982(25) &   .997(16) & 1.615(36) & 1.620(34) \\
0.115  &  .636(25) &   .663(13) & 1.097(26) & 1.105(24) \\
0.1225 &  .103(12) &   .130(12) &  .394(38) &  .403(37) \\
0.1240 &  .023(03) &   .031(04) &  .264(26) &  .266(26) \\ \hline
\end{array}$
\end{center}
\caption{
Hadron masses in the photon and the (matter)$\times$(monopole) fields
 on $16^3 \times 32$
 at $\beta=5.7$ and in the free fields.
}
\label{Tbl.ph-fr}
\end{table}

 \vspace{1cm}
 \vspace{1cm}

\end{document}